\documentclass[pre,a4paper,twocolumn,showpacs,amsmath]{revtex4}
\usepackage{graphicx,amsfonts,amsmath}

\begin{document}

\title{Indeterminacy, Memory, and Motion in a Simple Granular Packing}
\author{Sean McNamara}
\email[Email address:]{sean@ica1.uni-stuttgart.de}
\affiliation{Institut f\"ur Computerphysik, Universit\"at Stuttgart,
70569 Stuttgart, GERMANY}

\author{Ram\'on  Garc\'{\i}a-Rojo}
\affiliation{Institut f\"ur Computerphysik, Universit\"at Stuttgart,
70569 Stuttgart, GERMANY}

\author{Hans Herrmann}
\affiliation{Institut f\"ur Computerphysik, Universit\"at Stuttgart,
70569 Stuttgart, GERMANY}

\date{\today}

\begin{abstract}
We apply two theoretical and two numerical methods to the problem of
a disk placed in a groove and subjected to gravity and a torque.
Methods assuming
rigid particles are indeterminate -- certain combinations of forces
cannot be calculated, but only constrained by inequalities.  In methods
assuming deformable particles, these combinations of forces are determined
by the history of the packing.  Thus indeterminacy in rigid particles
becomes memory in deformable ones.  Furthermore, the torque needed to
rotate the particle was calculated.  Two different paths to motion were
identified.  In the first, contact forces change slowly,
and the indeterminacy decreases continuously to zero,
and vanishes precisely at the onset of motion, and the torque needed
to rotate the disk is independent of method and packing history.
In the second way, this torque depends on method and on the history
of the packing, and the forces jump discontinuously at the onset of motion.
\end{abstract}

\pacs{45.70.-n}
\maketitle

\section{Introduction}

\subsection{Motivation}

Static granular packings present many difficulties to theorists seeking
to understand them.  One complication is \textsl{indeterminacy}.
If the grains are assumed to be perfectly rigid but with friction, 
one cannot find a unique solution for the forces between them; there
are usually many possible solutions 
\cite{Radjai,FNensemble,Dietrich,us,ECOMASS}.
If one calculates forces from
particle displacements, indeterminacy disappears, but in many examples,
these displacements are tiny fractions of a grain diameter.  It is
astonishing that such tiny distances would change the fundamental
nature of the problem.  Furthermore, it is not clear how the unique
state, determined by particle displacements, is related to the set
of possible forces arising when the particles are rigid.  These
questions are made more pressing by the widespread use of simulation
methods that assume the particles are rigid, and hence somehow
choose one value for the contact forces out of many possible solutions.

Next, there is the question of \textsl{memory}.  The method of construction
of a packing and its history change its behavior.  For example, it has
been shown that the method of construction changes the distribution
of forces under a sand pile \cite{Vanel,Matuttis,Geng}.  In this case,
the memory of the packing seems to be stored in the orientation of
the contacts \cite{Geng}.

Finally, there is the onset of \textsl{motion}: what forces are needed to
permanently deform a static packing, and how does this motion originate?
These questions are the principal concern of soil mechanics.  Motion
often is very localized, i.e., in shear bands \cite{Oda1,Oda2}.  
Another phenomena related to the onset of motion is the linear
accumulation of deformation in granular assemblies subjected to
periodically varying forces \cite{Fernando,Ramon}.

In this paper, we consider a very simple system where the relationships
between these problems is clearly demonstrated.  We hope that it will
provide clues on how to understand much more complicate packings.

\subsection{Synopsis}

\begin{figure}
\begin{center}
\includegraphics[width=0.4\textwidth]{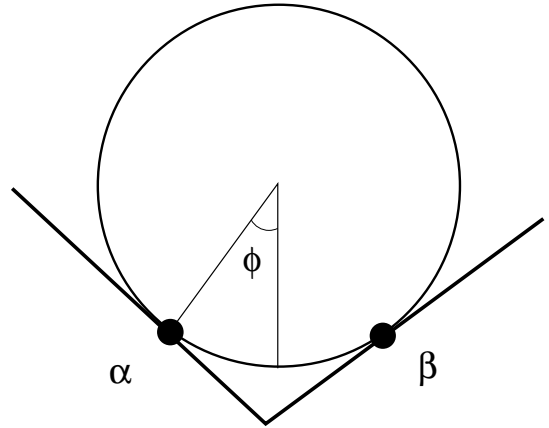}
\end{center}
\caption{\label{setup}A disk supported by two walls through
contacts $\alpha$ and $\beta$.  The angle $\phi$ suffices to characterize
the geometry.}
\end{figure}

The granular ``packing'' shown in Fig.~\ref{setup} is perhaps the
simplest possible granular packing.  Nevertheless, it exhibits many
subtleties, and in this paper we use it as an example to show the
relationship between indeterminacy, memory, and motion.  
This configuration is similar to the rod wedged between two
converging walls discussed in Ref.~\cite{ECOMASS}.
It is even
simpler than the one considered in Ref.~\cite{Halsey}, where a disk
is placed in an asymmetric groove and subjected to gravity.  Here,
we consider only a symmetric groove, but the forces we impose on
the disk are more complicated.  
In addition to a gravitational acceleration $g$, we apply a torque
$\tau>0$.  The contacts between the disk and the wall are assumed
to be non-cohesive and frictional, with Coulomb friction ratio $\mu$.
We investigate two questions:
How do the contact forces at $\alpha$ and $\beta$ change in response
to the imposed forces?  At what value of the torque does the particle begin
to rotate?

We first examine some general considerations: the equations of
static equilibrium, and the status of the contacts.
Then we attempt to deduce the contact forces from these considerations.
We show that for certain geometries and torques, one cannot decide whether
the disk rotates or not.  We then present a series of
contact dynamics (hereafter CD) \cite{CD}
and molecular dynamics (hereafter MD) \cite{MD}
simulations, and point out the difference between them.  
Then we formulate a second method of calculating the contact forces,
assuming the arise from small deformations of the particles.  This
second method explains all the features of the MD simulations,
and sheds light on the questions mentioned above.

\section{Rigid Particles}
\label{GeneralConsiderations}

\subsection{Static Equilibrium}

The equations of static equilibrium for the disk in Fig.~\ref{setup} are
\begin{eqnarray}
R_\alpha \sin\phi + T_\alpha \cos\phi - R_\beta \sin\phi
 + T_\beta \cos\phi &=& 0,\cr
R_\alpha \cos\phi - T_\alpha \sin\phi + R_\beta \cos\phi
 + T_\beta \sin\phi &=& mg,\cr
rT_\alpha + rT_\beta &=& -\tau.
\label{staticeq1}
\end{eqnarray}
Here, $R_\alpha$ indicates the normal force at contact $\alpha$,
with $R_\alpha>0$ for repulsion.  $T_\alpha$ is the tangential
force, with $T_\alpha>0$ when this force exerts a positive torque
on the disk.  $R_\beta$ and $T_\beta$ are defined in the same way.

These equations can be written in matrix form:
\begin{equation}
\mathbf{cF} + \mathbf{f}_\mathrm{ext}  = 0.
\label{staticeq2}
\end{equation}
We call $\mathbf{c}$ the \textsl{contact matrix}:
\begin{equation}
\mathbf{c} = \left( \begin{array}{cccc}
 \sin\phi & \cos\phi & -\sin\phi & \cos\phi \\
 \cos\phi & -\sin\phi & \cos\phi & \sin\phi \\
 0 & r & 0 & r
\end{array} \right).
\label{cmatrix}
\end{equation}
Note that the dimensions of $\mathbf{c}$ require that it have
at least one null eigenvalue.
The contact forces $\mathbf{F}$ and the external forces 
$\mathbf{f}_\mathrm{ext}$ are
\begin{equation}
\mathbf{F} = \left( \begin{array}c
	R_\alpha \\ T_\alpha \\ R_\beta \\ T_\beta
\end{array} \right), \quad
\mathbf{f}_\mathrm{ext} = \left( \begin{array}c
	0 \\ -mg \\ \tau 
\end{array} \right),
\label{vectorize}
\end{equation}
where $R_\alpha$ ($R_\beta$) are the normal components forces exerted at
contact $\alpha$ ($\beta$), and $T_\alpha$ and $T_\beta$ are the
tangential components.

It is convenient to use the following orthogonal basis
of $\mathbb{R}^4$:
\begin{eqnarray}
\mathbf{F}_x = \frac{1}{2} \left( \begin{array}c
	\sin\phi \\ \cos\phi \\ -\sin\phi \\ \cos\phi
\end{array} \right),\quad
\mathbf{F}_y = \frac{1}{2} \left( \begin{array}c
	\cos\phi \\ -\sin\phi \\ \cos\phi \\ \sin\phi
\end{array} \right),\cr
\mathbf{F}_\theta = \frac{1}{2} \left( \begin{array}c
	-\cos\phi \\ \sin\phi \\ \cos\phi \\ \sin\phi
\end{array} \right),\quad
\mathbf{F}_0 = \frac{1}{2}\left( \begin{array}c
	\sin\phi \\ \cos\phi \\ \sin\phi \\ -\cos\phi
\end{array} \right).
\label{CDvectors}
\end{eqnarray}
If $\mathbf{F}$ is written in this basis:
\begin{equation}
\mathbf{F} = a_x \mathbf{F}_x + a_y \mathbf{F}_y + a_\theta \mathbf{F}_\theta
 + a_0 \mathbf{F}_0,
\label{Fexpand}
\end{equation}
then one can verify that
\begin{equation}
\mathbf{cF} = \left( \begin{array}c
 a_x \\ a_y \\ a_x r \cos\phi + a_\theta r \sin\phi
\end{array} \right)
 = -\mathbf{f}_\mathrm{ext}.
\label{Fsolution}
\end{equation}
This equation enables one to easily construct solutions to 
Eq.~(\ref{staticeq2}), for one has immediately
$a_x=0$, $a_y=mg$, and $a_\theta=-\tau/(r\sin\phi)$.
However, $a_0$ cannot be
determined in this way because $\mathbf{F}_0$ is a null eigenvector
of $\mathbf{c}$.  This is an expression of the indeterminacy in the
problem.

\subsection{Contact Status}

Of course, one does not have absolute freedom to choose $a_0$,
because the following conditions must be satisfied at the contacts:
\begin{equation}
R \ge 0, \quad  \mu R \ge |T|,
\label{ContactConditions}
\end{equation}
where the constant $\mu$ is the Coulomb friction ratio.  The first
inequality excludes cohesive forces, and the second requires that the
tangential forces not exceed a certain threshold.  

At this point, it is useful to introduce the concept of contact
\textsl{status}.  If we have equality in the second condition
in Eq.~(\ref{ContactConditions}), the contact is said to be
\textsl{sliding}, otherwise it is \textsl{non-sliding}. 
The tangential relative motion $v_t$ must be consistent with the
contact status.  If the contact is non-sliding, no tangential motion
is allowed, but if the contact is
sliding, then the tangential force must oppose the motion.  Note that
$v_t=0$ is allowed under all circumstances, so that contacts in
static equilibrium can be sliding.

To see how this limits the possible values of $a_0$, one can first
use the definition of $\mathbf{F}$ in Eq.~(\ref{vectorize}) and
the basis in Eq.~(\ref{CDvectors}) to compute the forces in terms
of $\mu$, $\phi$, and the coefficients in the expansion 
of Eq.~(\ref{Fexpand}).  Then one can use Eq.~(\ref{Fsolution}) to
eliminate all the coefficients except for $a_0$.  
Finally, one forms the inequalities $\mu R_\alpha \ge -T_\alpha$,
$\mu R_\alpha \ge T_\alpha$, $\mu R_\beta \ge -T_\beta$,
and $\mu R_\beta \ge T_\beta$.  These yield
\begin{subequations}
\begin{eqnarray}
a_0
\left(\mu\tan\phi+1\right) +
\left[ mg + \frac{\tau}{r\sin\phi} \right]
\left(\mu-\tan\phi\right) \ge 0,
\label{alpha+}\\
a_0
\left(\mu\tan\phi-1\right) +
\left[ mg + \frac{\tau}{r\sin\phi} \right]
\left(\mu+\tan\phi\right) \ge 0,
\label{alpha-}\\
a_0
\left(\mu\tan\phi-1\right) +
\left[ mg - \frac{\tau}{r\sin\phi} \right]
\left(\mu+\tan\phi\right) \ge 0,
\label{beta+}\\
a_0
\left(\mu\tan\phi+1\right) +
\left[ mg - \frac{\tau}{r\sin\phi} \right]
\left(\mu-\tan\phi\right) \ge 0,
\label{beta-}
\end{eqnarray}
\label{allconditions}
\end{subequations}
Some of these conditions are redundant.  For example,
Eq.~(\ref{beta+}) implies Eq.~(\ref{alpha-}).  To see this,
note that due to our choice $\tau>0$,
\begin{equation}
\left[ \frac{2\tau}{r\sin\phi} \right] 
\left(\mu+\tan\phi\right) \ge 0.
\end{equation}
Adding this to Eq.~(\ref{beta+}) gives Eq.~(\ref{alpha-}).  Thus
Eq.~(\ref{alpha-}) never needs to be considered; it will be satisfied
as long as Eq.~(\ref{beta+}) is.  Similar reasoning can be used to
show that Eq.~(\ref{alpha+}) is redundant when $\mu-\tan\phi>0$,
and that Eq.~(\ref{beta-}) is redundant when $\mu-\tan\phi<0$.
When $\mu=\tan\phi$, these conditions are equivalent.

The conditions in Eqs.~(\ref{allconditions}) put limits on the value that
$a_0$ can attain \cite{us}.  Furthermore, when a contact is sliding,
we have equality in one of the above cases, and $a_0$ is determined:
indeterminacy disappears.

\subsection{Application}
\label{CD}

We now attempt to deduce the contact forces and
particle motion, using only the considerations presented above.
No relation between particle deformation and contact forces is
assumed.  Note that the biggest difficulty will be determining
the coefficient $a_0$; the other three coefficients in the 
expansion in Eq.~(\ref{Fexpand}) can be read off from 
Eq.~(\ref{Fsolution}).  The unknown value of $a_0$ is an
expression of the indeterminacy of the problem.  We must consider
three separate cases, depending on the relation of $\mu$
to the slope of the sides of the groove, $\tan\phi$.

\subsubsection{Shallow slope: $\tan\phi<\mu$}
\label{CDshallow}

Let us first restrict ourselves to $\tan\phi<\mu$ and $\mu<1$.
Under these assumptions, we have $\mu-\tan\phi>0$, so the $a_0$ is
constrained by the conditions Eqs.~(\ref{beta+}) and (\ref{beta-}).
Furthermore, $\mu\tan\phi-1<0$, so that
Eq.~(\ref{beta+}) sets a lower, and Eq.~(\ref{beta-})
an upper bound on $a_0$:
\begin{equation}
a_\mathrm{min} \le a_0 \le a_\mathrm{max},
\label{aminmax}
\end{equation}
where
\begin{eqnarray}
a_\mathrm{min} = -\left[ mg - \frac{\tau}{r\sin\phi} \right]
  \left( \frac{\mu-\tan\phi}{1+\mu\tan\phi} \right),\cr
a_\mathrm{max} = \left[ mg - \frac{\tau}{r\sin\phi} \right]
  \left( \frac{\mu+\tan\phi}{1-\mu\tan\phi} \right).
\label{aminmaxShallow}
\end{eqnarray}
The quantities in the curved brackets are always positive, while
the quantity in the square brackets
is positive for $\tau=0$, and decreases as $\tau$ increases.
It vanishes when
\begin{equation}
\tau = \tau_1 \equiv mgr \sin\phi,
\end{equation}
and becomes negative when $\tau>\tau_1$.  When it is negative, no
solution for $a_0$ exists, for $a_\mathrm{min} > a_\mathrm{max}$.
Thus there are no solutions of the equations of static equilibrium,
when $\tau>\tau_1$ that also satisfy Eq.~(\ref{ContactConditions}), and
the disk must move.  On the other hand,
when $\tau<\tau_1$, an infinite number of such solutions exist, 
one for each value of $a_0$ satisfying 
Eq.~(\ref{aminmax}).  Finally, at $\tau=\tau_1$,
we have $a_\mathrm{min}=a_0=a_\mathrm{max}=0$, hence there is a unique
solution.

We have shown that the disk must rotate when $\tau>\tau_1$, and that
static solutions exist for $\tau\le\tau_1$, but we have not yet shown
that the disk will not move when $\tau\le\tau_1$.  After all, for
some choice of $a_0$, it might be possible to put the disk in motion.
But this can be shown to be impossible by considering what happens at
$\tau=\tau_1$.  In this situation, we have $a_0=0$, so the contact
forces are
\begin{equation}
R_\alpha = mg \cos\phi, \quad T_\alpha = - mg \sin\phi, \quad
R_\beta = T_\beta = 0.
\label{rolling}
\end{equation}
Thus the disk is entirely supported by contact $\alpha$.  $R_\alpha$
and $T_\alpha$ cancel the gravitational force, and $\tau$ balances
the torque $rT_\alpha=-\tau_1$ exerted on the disk by the contact force.
A small increase in $\tau$ and $T_\alpha$ will cause the disk to
roll upwards to the left, out of the groove.  On the other hand,
if $\tau<\tau_1$, then the torque $rT_\alpha=-\tau_1$ will not be
balanced, and the disk will try to roll down the slope.  Therefore,
if $\tau<\tau_1$, the disk cannot rotate.

Another possibility is that the disk rotate in place.  But this is
impossible, when $\tan\phi<\mu$, because this requires
$T_\alpha=-\mu R_\alpha$ and $T_\beta=-\mu R_\beta$, i.e., contacts
$\alpha$ and $\beta$ must be sliding.  This occurs when we have
equality in both Eqs.~(\ref{alpha+}) and (\ref{beta+}).  But this
cannot happen because when equality holds in Eq.~(\ref{alpha+}),
then Eq.~(\ref{beta-}) is violated.

Therefore, the value of $a_0$ has no effect on the motion of the disk
If $\tau<\tau_1$,
the disk remains in the groove, although the forces cannot be uniquely
determined.  At $\tau=\tau_1$, there is a unique solution for the forces:
contact $\beta$ opens and the particle's weight is supported by contact
$\alpha$.  For $\tau>\tau_1$, the disk rolls out of the groove.

Note that there is a relationship between the onset of motion and the
disappearance of indeterminacy: we have motion for $\tau>\tau_1$ and
indeterminacy for $\tau<\tau_1$.  Indeed, we could have anticipated
this when we wrote down Eqs.~(\ref{allconditions}).  For a disk to
move, at least one contact must be sliding (or open), leading to
an equality in at least one of Eqs.~(\ref{allconditions}).  Then
this equality can be used to determine $a_0$, eliminating the
indeterminacy.

\subsubsection{Intermediate slope: $\mu < \tan\phi < 1/\mu$}

Now let us consider intermediate angles $\mu\le\tan\phi\le1/\mu$.
At these values of $\phi$, the relevant conditions from 
Eqs.~(\ref{allconditions}) are Eqs.~(\ref{alpha+}) and (\ref{beta+}).
Eq.~(\ref{beta+}) sets an upper, and Eq.~(\ref{alpha+}) a lower,
bound on $a_0$.  Thus we have Eq.~(\ref{aminmax}) again, except now
\begin{eqnarray}
a_\mathrm{min} = \left[mg + \frac{\tau}{r\sin\phi} \right]
\left( \frac{\tan\phi-\mu}{1+\mu\tan\phi} \right),\cr
a_\mathrm{max} = \left[mg - \frac{\tau}{r\sin\phi} \right]
\left( \frac{\tan\phi+\mu}{1-\mu\tan\phi} \right).
\label{aminmaxIntermediate}
\end{eqnarray}
Again, the quantities in the curved brackets is positive for
$\mu < \tan\phi < 1/\mu$.  The quantities in the square brackets
are equal and positive for $\tau=0$.  But as $\tau$ increases,
$a_\mathrm{min}$ increases while $a_\mathrm{max}$ decreases.
When
\begin{equation}
\tau = \tau_2 \equiv 
   mgr \frac{\mu}{1+\mu^2}\sec\phi,
\label{taumax2}
\end{equation}
we have $a_\mathrm{min} = a_0 = a_\mathrm{max} = a_2$, where
\begin{equation}
a_2 =
mg \left[1 + \frac{\mu}{\tan\phi}\frac{1+\tan^2\phi}{1+\mu^2}\right]
\left( \frac{\tan\phi-\mu}{1+\mu\tan\phi}\right).
\label{aslip2}
\end{equation}
and there is a unique
solution for the contact forces.  When $\tau>\tau_2$, no static
solution satisfying the contact conditions exists, as
$a_\mathrm{min} > a_\mathrm{max}$.

Furthermore, we can show that the disk cannot move when $\tau<\tau_2$.
The disk cannot roll out of the groove, because
the forces in Eq.~(\ref{rolling}) violate $|T_\alpha|\le\mu R_\alpha$.
If the disk is to rotate in place, we must have
$T_\alpha = -\mu R_\alpha$ and $T_\beta = -\mu R_\beta$, i.e., we must
have equality in Eqs.~(\ref{alpha+}) and (\ref{beta+}).  But we have
just showed that requiring these equalities leads to $\tau=\tau_2$, 
$a_\mathrm{min}=a_0=a_\mathrm{max}$.  

Therefore, we have exactly the same situation as in Sec.~\ref{CDshallow}.
For $\tau<\tau_2$, there is indeterminacy without motion, and for
$\tau>\tau_2$ there is motion.  There is a unique solution
for the forces precisely at $\tau=\tau_2$.  Furthermore, note
that $\tau_1 = \tau_2$ when $\tan\phi=\mu$ and when $\tan\phi=1/\mu$.

\subsubsection{Large slopes: $\tan\phi>1/\mu$}

At large angles ($\tan\phi>1/\mu$), the relevant conditions are
still Eqs.~(\ref{alpha+}) and (\ref{beta+}) as in the previous
section.  But now the factor $1-\mu\tan\phi$ becomes negative,
transforming Eq.~(\ref{beta+}) into a \textsl{lower} bound on $a_0$.
Eq.~(\ref{alpha+}) remains a lower bound on $a_0$, meaning that
there is no upper bound on $a_0$.  It
can be made arbitrarily large.  Therefore, for any $\tau>0$, it is
possible to satisfy all conditions in Eq.~(\ref{allconditions}) simply
by making $a_0$ very large.  (Note that all the factors multiplying
$a_0$ in these conditions are positive when $\tan\phi>1/\mu$).
This means that an infinite torque can be
put on the disk without causing it to rotate.

On the other hand,
if $\tau=\tau_2$, one can obtain equality in Eqs.~(\ref{alpha+})
and (\ref{beta+}), meaning contacts $\alpha$ and $\beta$ are sliding.
If $\tau$ is increased slightly, then the disk can start to rotate
in place.  Therefore, both static and moving solutions co-exist
for $\tau>\tau_2$, and we cannot decide on the basis of
Eqs.~(\ref{allconditions}) whether the disk rotates or not.

This remarkable situation poses two questions:  First, how will
the CD algorithm handle this case?  This algorithm tries to use the
information presented in Sec.~\ref{GeneralConsiderations}
to deduce the motion of the particles, but we have shown that
this is insufficient.  Secondly, it would be very easy to
construct an experiment to measure the torque needed to make the disk
rotate.  One could construct a groove with $\tan\phi>1/\mu$, and
place a cylinder in the groove, and try to turn the cylinder.  What
determines whether the cylinder rotate?  These questions are
investigated in the next sections.

\section{Simulations}

\begin{figure}
\begin{center}
\includegraphics[width=0.48\textwidth]{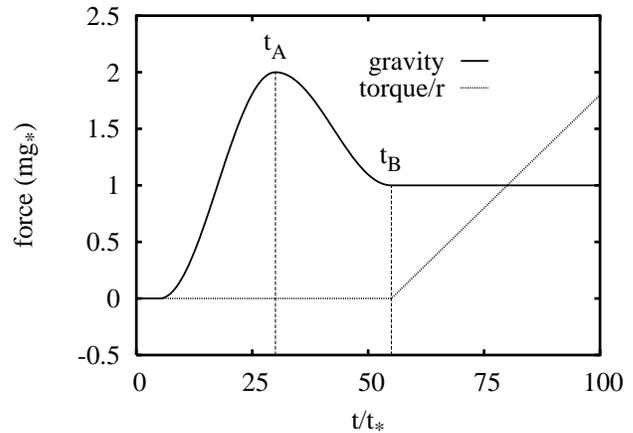}
\end{center}
\caption{\label{forces}The forces applied to the disk.  Time is given
in units of $t_*=\sqrt{d/g}$.  The gravity attains a maximum at
$t=t_A$, and the torque begins to increase at $t=t_B$.}
\end{figure}

To verify the conclusions of the previous section, and to clarify
the situation for $\tan\phi>1/\mu$, we carried out CD and MD simulations.
The disk was placed in the groove without any forces acting on it.
Then gravity was turned on, and increased, until it reached a maximum of
$g_\mathrm{max}$ at $t=t_A$.  Then it was decreased,
reaching $g_*$ at $t=t_B$, and thereafter was held constant.  
A linearly increasing torque $\tau$ was then applied
until the particle began to rotate.  Fig.~\ref{forces} shows the 
gravity force and torque as functions of time.  In the following,
we will vary $g_\mathrm{max}$, but keep $g_*$ fixed.  
In all cases, $\mu=0.5$.

The experiment can be considered as a very simple soil mechanics
experiment.  First, the sample, consisting of one disk,
is ``prepared'' by pushing the disk into the groove,
and ``loaded'' by the torque until it ``yields''.
We are especially interested in knowing if and how the preparation
affects the yielding torque.

\begin{figure}
\begin{center}
\includegraphics[width=0.48\textwidth]{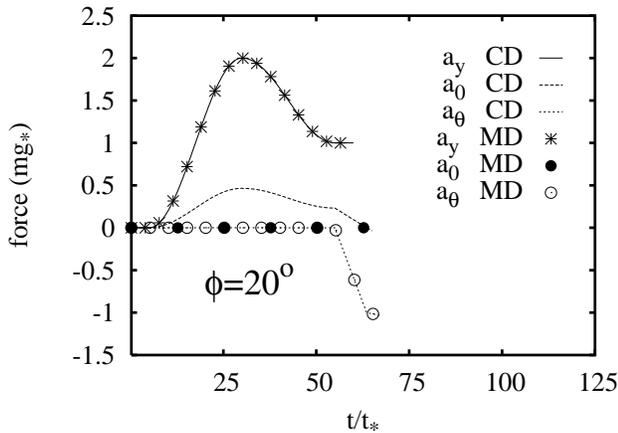}
\end{center}
\caption{\label{small} The coefficients of the expansion,
Eq.~(\ref{Fexpand}), as a function of time.  Here $\phi=20^\circ$
and $\mu=0.5$ so that $\tan\phi<\mu$.  The curves stop when the
particle begins to rotate.} 
\end{figure}

A typical result of a simulation is shown in Fig.~\ref{small}.
The three non-zero coefficients $a_y$, $a_\theta$, and $a_0$
of the expansion in Eq.~(\ref{Fexpand}) are shown.  In accordance
with Eq.~(\ref{Fsolution}), $a_y$ and $a_\theta$ are identical
for the CD and MD solution methods, and given by
$a_y = mg$, $a_\theta = -\tau/(r\sin\phi)$.  The coefficient
$a_0$, however, differs between the CD and MD solutions.
In the following, we will plot only $a_0$, as the other coefficients
are always uniquely determined by the imposed forces.  
Three different values of $\phi$ will be considered, corresponding
to the three subsections of Sec.~\ref{CD}.

\subsection{Shallow Slopes: $\tan\phi<\mu$.}
\begin{figure}
\begin{center}
\includegraphics[width=0.48\textwidth]{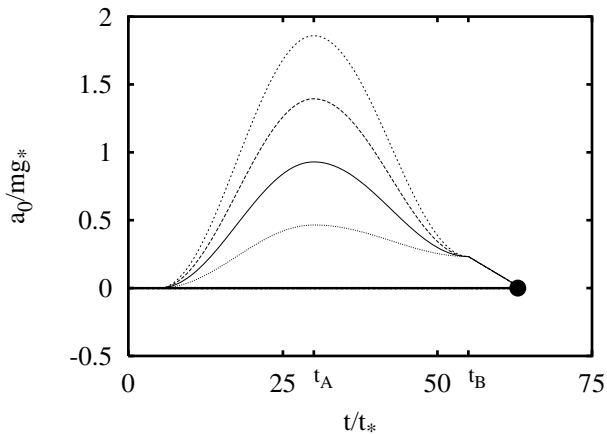}
\end{center}
\caption{\label{smallscan} The coefficient $a_0$ as a function
of time for various values of
$g_\mathrm{max}/g_*$.  Here $\phi=20^\circ$ and $\mu=0.5$, 
so that $\tan\phi<\mu$.
The thin lines are the CD simulations, with $g_\mathrm{max}/g=2$ 
(lowest curve), $4$, $6$, and $8$ (highest curve).
The thick line are the corresponding MD simulations, which fall
on top of each other in this case.  The large dot indicates the point
where the disk begins to rotate.}
\end{figure}

In Fig.~\ref{smallscan}, we plot $a_0$ as a function of time for the
MD and CD simulations.  In the
MD simulations, $a_0=0$, independent of time and of $g_\mathrm{max}$.  
In the CD simulation, $a_0$ is proportional to the gravity for $t<t_B$.
Note that the values of $a_0$ are
significant -- almost twice the weight of the particle for
$g_\mathrm{max}/g_*=8$.
However, at $t=t_B$, all CD simulations have the same value of $a_0$.
As the torque is applied, $a_0$ moves
linearly towards $0$.  This occurs because when $\tau\le \tau_1$,
the disk cannot move, but $a_0$ is constrained between the two values
given in Eq.~(\ref{aminmaxShallow}),
which approach each other as $\tau$ approaches $\tau_1$.  
Indeed, the diagonal line segment near $50 < t/t_* < 60$ traced
out by the CD simulations corresponds to requiring equality
in Eq.~(\ref{beta-}).  The conditions Eqs.~(\ref{beta+}) and
(\ref{beta-})
act as a ``funnel'' that guides $a_0$ toward $0$ as
the torque increases, so that $a_0=0$ when $\tau=\tau_1$.  Thus
the MD and CD both predict that the disk moves at $\tau=\tau_1$.
However, the two methods predict different routes to failure.
MD predicts that both contacts remain non-sliding until the
force at $\beta$ vanishes, and the disk starts to roll.  On the
other hand, CD predicts that the contact at $\beta$ first becomes
sliding, and then later starts to roll.

\subsection{Intermediate Slopes: $\mu < \tan\phi < 1/\mu$.}
\begin{figure}
\begin{center}
\includegraphics[width=0.48\textwidth]{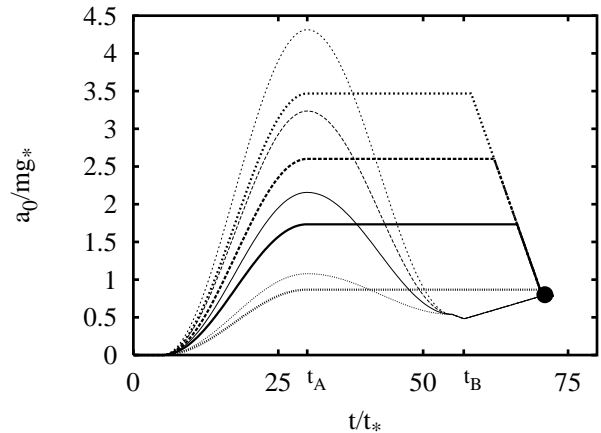}
\end{center}
\caption{\label{intermediate}The undetermined coefficient
$a_0$ from Eq.~(\ref{Fexpand}) as a function of time.
Here, $\phi=50^\circ$ and $\mu=0.5$,
so that $\mu<\tan\phi<1/\mu$. 
The thin lines are the CD simulations, with $g_\mathrm{max}/g=2$ 
(lowest curve), $4$, $6$, and $8$ (highest curve).
The thick lines are the MD simulations with the same values of
$g_\mathrm{max}/g$.  The large dot indicates the point
where the disk begins to rotate.}
\end{figure}

In Fig.~\ref{intermediate}, we show $a_0$ for the case of
intermediate slope $\mu < \tan\phi < 1/\mu$.  
The behavior of the CD simulation is quite similar to the
preceding case.  For $t<t_B$, $a_0$
is again proportional to gravity, but attaining even larger values
than before.  At $t=t_B$, all CD simulations
are identical, and they evolve together during the loading.  They
again encounter the ``funnel'' arising from the lower and upper
bounds on $a_0$ given in Eq.~(\ref{aminmaxIntermediate}), and
move toward $a_0=a_2$, and the disk starts to rotate when
$\tau=\tau_2$.

The MD simulation has a quite different
behavior.  For $t<t_A$, $a_0$ is also proportional to the gravity,
although lower than the CD value.  But for $t>t_A$, $a_0$ remains
constant, so that $a_0$ is proportional to $g_\mathrm{max}$, even
at $t=t_B$.  In this
case, therefore, the preparation does make a difference.
The MD simulations remember the value of $g_\mathrm{max}$, 
and this memory is stored in $a_0$.
This simple example clearly demonstrates a principle stated
in the paper's introduction: the indeterminacy of forces in
perfectly rigid particles is related to history-dependency in
packings of deformable particles.  The coefficient $a_0$, which
is undetermined in the perfectly rigid case, here contains
the memory of the packing, and is simply proportional to the
greatest downward force that the disk has experienced in the past.

The history dependency of the MD simulation, however, does not
affect the yielding torque.  The different MD simulations encounter
the funnel at different times, but are all guided towards the 
point $\tau=\tau_2$,
$a_0=a_2$, where the disk begins to rotate.  Therefore, the
memory on the MD simulation can only be detected by inspecting
the contact forces.  

\subsection{Large Slopes: $\tan\phi>1/\mu$.}
\begin{figure}
\begin{center}
\includegraphics[width=0.48\textwidth]{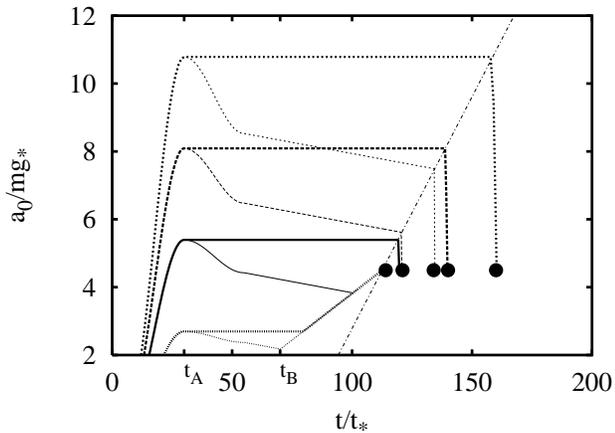}
\end{center}
\caption{\label{steep}Coefficients of the expansion 
Eq.~(\ref{Fexpand}) as a function of time.  Here, $\phi=80^\circ$
and $\mu=0.5$, so that $\tan\phi>1/\mu$. 
The thin lines are the CD simulations, with $g_\mathrm{max}/g=2$ 
(lowest curve), $4$, $6$, and $8$ (highest curve).
The thick lines are the MD simulations with the same values of
$g_\mathrm{max}/g$.  The dots indicates the points
where the disk begins to rotate.
The diagonal dot-dashed line is obtained by requiring equality in
Eq.~(\ref{beta+}).
}
\end{figure}

Finally, in Fig.~\ref{steep}, we show the case where $\tan\phi>1/\mu$.
This case has some important differences from the preceding cases.
First of all, both the CD as well as MD exhibits history-dependence, as the
contact forces at the end of the sample preparation and the
yielding torque depend on $g_\mathrm{max}$.
Secondly, only half of the funnel observed
at lower angles operates as before.

One may ask how the CD simulation can exhibit memory, because it
deduces the contact forces from the principles stated in
Sec.~\ref{GeneralConsiderations}, but the history of the packing
does not enter into those considerations.  The memory of the CD
algorithm comes from the way it chooses the contact forces.  
It begins with an initial guess, which it then refines through an
iterative process until it arrives at a satisfactory solution.
One usually takes the initial guess to be the solution of the previous
time step because convergence is faster.  But in Fig.~\ref{steep}, it is
clear that this choice of initial guess also serves to make the
CD algorithm history dependent, as the solution chosen depends on
the initial guess \cite{ECOMASS}. 
But this memory is not the same as in the MD
simulations.  Nor is it clear why the preparation of the sample
makes a difference at $\phi=80^\circ$ but not at $\phi=50^\circ$.

But the most important difference with the previous examples is that
only half of the funnel works as before.  As was noted earlier, there
is no longer an upper bound on $a_0$; Eqs.~(\ref{alpha+}) and
(\ref{beta+}) are both lower bounds.  From the figure, one can see
that the lower bound set by Eq.~(\ref{alpha+}) functions as before.
When $a_0$ reaches this lower bound, it then increases linearly as
the torque is increased, until equality is obtained in both
Eqs.~(\ref{alpha+}) and (\ref{beta+}).  Then the disk begins to
rotate.  The behavior of the lower bound set by Eq.~(\ref{beta+})
is quite different.  When equality is obtained in that
condition, the value of $a_0$ jumps discontinuously, and the
particle begins to rotate.  As a result, the yielding torque
is determined by when the system first satisfies
equality in Eq.~(\ref{beta+}), and this in turn depends on the
value of $a_0$ set by the preparation phase of the experiment.
Thus in this case, the memory does affect the yielding torque.
In Sec.~\ref{MD}, we will explain this discontinuity in $a_0$
theoretically.  But one can already understand its physical
origin.  Equality in Eq.~(\ref{beta+}) corresponds to contact
$\beta$ becoming sliding, meaning that tangential motion is
now possible there.  Since contact $\alpha$ is non-sliding, the
particle pivots about contact $\alpha$.  As we showed earlier,
contact $\beta$ cannot open when $\tan\phi>\mu$, but 
the disk can apparently move enough to reduce
the contact forces very quickly, allowing the particle to rotate.

\begin{figure}
\begin{center}
\includegraphics[width=0.48\textwidth]{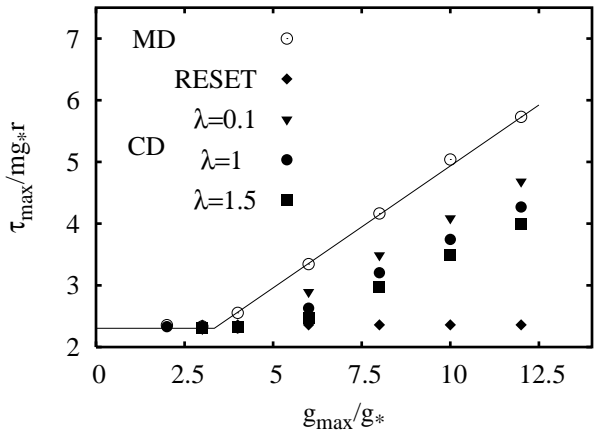}
\end{center}
\caption{\label{scan}The torque at which the disk begins to rotate at 
$\phi=80^\circ$ and $\mu=0.5$, 
for different values of $g_\mathrm{max}$.  The four
different sets of points for CD correspond to the different iteration
algorithms described in the text.  The line shows the theoretical
prediction given in Sec.~\ref{steepMDtheory}.}
\end{figure}
In Fig.~\ref{scan}, we show the yielding torque as a function of
$g_\mathrm{max}$.  It is initially
independent of $g_\mathrm{max}$, corresponding to the case where the
system first meets the lower edge of the funnel.
Then, CD and MD give different yielding torques.  The MD values are well
predicted by a result that will be obtained in Sec.~\ref{steepMDtheory}.

The CD values for the yielding torque are equal to or below the MD ones,
because $a_0$ decreases as the torque is increased, while in the
MD case, $a_0$ remains constant.  This is plainly visible in Fig.~\ref{steep}.
The different CD points correspond to different iteration procedures.
The CD algorithm searches for a solution by adjusting the contact
forces one by one.  It considers a given contact, and calculates the
change in its contact forces needed to prevent interpenetration
and minimize sliding.  It then adds this to change to the its
current guess for the forces.  After passing over all the contacts
a certain number of times, the changes in the force needed fall below
a certain threshold, and the solution is accepted.  But this procedure
can be altered by multiplying the calculated change in forces by a 
number $\lambda$, with $\lambda=1$ corresponding to the usual iteration
procedure.  Changing $\lambda$ changes the yielding torque, as shown
above.  Furthermore, the memory in the CD algorithm can be erased at each
time step by always using $\mathbf{F}=0$ as the initial guess, instead of the
solution of the last time step.  When this is done, one obtains the points
labelled ``RESET'' in Fig.~\ref{scan} and the yielding torque
is independent of $g_\mathrm{max}$.  This result emphasizes the important role
of memory in this experiment.

\section{Deformable Particles}

\subsection{Formalism}

We now present a second method of calculating the contact forces
that is able to explain all features of the MD simulations
presented in the previous section.  The cost of
this additional information is that more assumptions must be made.  
First of all, one must specify how the
forces depend on the deformations, but more significantly, one
must specify the past history of the packing.  
This method shows how \textsl{indeterminacy} is replaced by \textsl{memory}.
More precisely, when $\tan\phi>\mu$, we show that
$a_0$ in Eq.~(\ref{Fexpand}) stores information about the largest
downward force that has been exerted on the disk in the past.
Indeed, $a_0$ is a quantification of the ``degree of wedging'' discussed in
Ref.~\cite{ECOMASS}.

We model
the deformations in a very simple way, assuming that contact forces are
generated by springs that are stretched by the motion of the disk.  
Our model is inspired by the MD simulation method
\cite{MD}, but we apply it analytically to our simple problem.  
A more general and realistic calculation has been
presented before \cite{Halsey}, but our purpose here is to show how
the indeterminacy is replaced by history-dependence when forces arise 
from deformations.

When two bodies touch, 
a normal and a tangential spring be created at the instant of contact. 
The contact forces are simply proportional to the spring lengths:
\begin{equation}
R = - k \delta_n, \quad T = - k \delta_t,
\label{Fscalar}
\end{equation}
where $\delta_n$ and $\delta_t$ are the normal and tangential spring
lengths, and $k$ is the spring constant.  
One normally includes damping in Eq.~(\ref{Fscalar}), but we will
assume that the motion is quasi-static, i.e., a sequence of equilibrium
states.  Under this assumption, the particle velocities are vanishingly
small, and so are the damping forces.  However, the damping has
important consequences that will be discussed below.
Eq.~(\ref{Fscalar}) leads to
a vector equation for $\mathbf{F}$ in terms of the spring lengths
$\mathbf{D}$:
\begin{equation}
\mathbf{F} = -k \mathbf{D},
\label{Fmat}
\end{equation}
where $\delta_{n,\alpha}$, $\delta_{t,\alpha}$, $\delta_{n,\beta}$,
and $\delta_{t,\beta}$ are gathered into $\mathbf{D}$ just as the
contact forces are arranged in $\mathbf{F}$ [see Eq.~(\ref{vectorize})].

In general, one
could introduce different spring constants for the normal and tangential
springs, and the ratio of these spring constants affects the behavior
of the system, even in the limit of infinitely hard particles.  
We set both spring constants equal, because it simplifies
the results, and our purpose here is only to show how the indeterminacy
is removed.  But we remark that the
ratio of the spring constants does not appear when the particles
are rigid, indicating that the physics associated with this parameter has
been eliminated.

The spring lengths obey
\begin{equation}
\dot\delta_n = v_n, \quad
\dot\delta_t = \left\{ \begin{array}{l}
v_t,\\
\pm\mu v_n,
\end{array}\right.
\label{Dscalar}
\end{equation}
where $v_n$ and $v_t$ are the normal and tangential components of the
relative velocity.  There are two choices for $\dot\delta_t$ because
$\mu R \ge |T|$ means that the tangential spring has a maximum
allowable length: $|\delta_t| \le \mu \delta_n$.
If applying $\dot\delta_t = v_t$ would lead to a violation of this
condition, the second choice is taken. 
Eq.~(\ref{Dscalar}) relates the time derivative of $\mathbf{D}$ to the
velocity of the disk.  If all contacts are non-sliding, we have
\begin{eqnarray}
\dot\delta_{n,\alpha} &=& v_x \sin\phi + v_y \cos\phi,\cr
\dot\delta_{t,\alpha} &=& v_x \cos\phi - v_y \sin\phi + r\omega,\cr
\dot\delta_{n,\beta} &=& -v_x \sin\phi + v_y \sin\phi,\cr
\dot\delta_{t,\beta} &=& v_x \cos\phi  + v_y \sin\phi + r\omega.
\label{stretching1}
\end{eqnarray}
Gathering the velocities of the disk into a single vector:
\begin{equation}
\mathbf{v} = \left( \begin{array}c v_x \\ v_y \\ \omega \end{array} \right),
\end{equation}
Eqs.~(\ref{stretching1}) becomes
\begin{equation}
\dot{\mathbf{D}} = \mathbf{c}^T \mathbf{v},
\label{stretching2}
\end{equation}
where $\mathbf{c}^T$ is the transpose of $\mathbf{c}$ in Eq.~(\ref{cmatrix}).
The appearance of $\mathbf{c}^T$ is not a coincidence, but a general
property valid for all granular packings \cite{Roux}.

It remains to incorporate the status of the contacts into our formalism.
This can be be done by inserting a matrix $\mathbf{S}$ that depends
on the contact status into Eq.~(\ref{stretching2}):
\begin{equation}
\dot{\mathbf{D}} = \mathbf{Sc}^T \mathbf{v},
\label{stretching3}
\end{equation}
where
\begin{equation}
\mathbf{S} = \left( \begin{array}{cc}
  S_\alpha & 0 \\ 0 & S_\beta
\end{array} \right). 
\end{equation}
Here, $S_\alpha=1$ if contact $\alpha$ is non-sliding, $S_\alpha=0$
if it is open, and 
\begin{equation}
S_\alpha =  \left( \begin{array}{cc}
1 & 0 \\ \pm\mu & 0
\end{array} \right),
\end{equation}
if it is sliding.

Finally, let us not forget Newton's equation of motion:
\begin{equation}
\mathbf{M}\dot{\mathbf{v}} = \mathbf{cF} + \mathbf{f}_\mathrm{ext}.
\label{Newton}
\end{equation}
Here, the matrix $\mathbf{M}$ contains the mass $m$ and moment
of inertia $I$ of the disk on the diagonal:
\begin{equation}
\mathbf{M} = \left( \begin{array}{ccc}
m & 0 & 0\\ 0 & m & 0 \\ 0 & 0 & I
\end{array} \right).
\end{equation}

Eqs.~(\ref{Fmat}), (\ref{stretching3}), and (\ref{Newton}) form a set
of equations that can be solved for the motion of the disk.
Note that $\mathbf{c}$ remains constant because all contacts are
between the disk and the straight walls.  
The status of the contacts can change, however, so it is useful to
divide time up into segments $[t_0,t_1]$, $[t_1,t_2], \ldots$,
where the changes of contact status occur at the times
$t_1$, $t_2,\ldots$.  In the interior of a time interval,
the matrices $\mathbf{c}$, $\mathbf{S}$, and $\mathbf{M}$ are constant,
so Eqs.~(\ref{Fmat}), (\ref{stretching3}), and (\ref{Newton}) can
be combined into
\begin{equation}
\mathbf{M}\ddot{\mathbf{v}} = -\mathbf{Qv} + \dot{\mathbf{f}}_\mathrm{ext},
\label{Qmat}
\end{equation}
where $\mathbf{Q} = k\mathbf{c}\mathbf{Sc}^T$.
The matrix $\mathbf{Q}$ gives the change
in force that arises from a small displacement of the disk.  
From this equation, the importance of the sign of
$\mathbf{v}^T\mathbf{Qv}$ can be seen.
If $\mathbf{v}^T\mathbf{Qv}>0$,
then $\mathbf{Q}$ acts like a nonnegative number in Eq.~(\ref{Qmat}),
so that the contacts generate forces opposing the
velocities $\mathbf{v}$, and hence 
$\mathbf{v}$ will tend to oscillate about an equilibrium value
depending on $\dot{\mathbf{f}}_\mathrm{ext}$.
If the damping is chosen properly, these oscillations will be damped out
very quickly, so $\mathbf{v}$ will always be close to its equilibrium
value.  Physically, this means that $\mathbf{v}$ is slaved to 
$\dot{\mathbf{f}}_\mathrm{ext}$, as long as $\mathbf{f}_\mathrm{ext}$
changes much more slowly than the damping time scale.
This amounts to removing the inertia from Eq.~(\ref{Qmat}) and obtaining
\begin{equation}
\mathbf{Qv} = \dot{\mathbf{f}}_\mathrm{ext}.
\end{equation}
The situation is different if $\mathbf{f}_\mathrm{ext}$ causes
displacements that are in the null space of $\mathbf{Q}$,
i.e., $\mathbf{v}^T\mathbf{Qv}=0$.  In this
case, no forces opposing the displacements are generated, and they
can grow without bound, leading to permanent re-arrangements of the
particles.  Another possibility
is that $\mathbf{v}^T\mathbf{Qv}<0$.  In this case, the contact
forces amplify the velocities, which then grow exponentially.  In
this case, the packing fails catastrophically, with the contact
forces changing much more quickly than $\mathbf{f}_\mathrm{ext}$.

Finally, let us note there is a compatibility condition between
$\mathbf{S}$ and $\mathbf{v}$ that must be satisfied: $\mathbf{v}$
must be such that the tangential springs stay stretched to their
maximum lengths.  If this condition is not fulfilled at a contact,
then that contact becomes non-sliding, and $\mathbf{S}$ must be
modified.

Let us now show how all this can be used to calculate the contact
forces without indeterminacy.  We will do so by calculating the
coefficients of the expansion in Eq.~(\ref{Fexpand}).  We can
construct an equation for these coefficients by differentiating
Eq.~(\ref{Fmat}) with respect to time, and combining it with
Eq.~(\ref{stretching3}), leading to 
$\dot{\mathbf{F}} = \mathbf{Sc}^T\mathbf{v}$.  Then we project
this equation onto the basis given in Eq.~(\ref{CDvectors})
by left-multiplying it by 
$\mathbf{a}$, a $4\times4$ matrix whose rows are the basis
vectors in Eq.~(\ref{CDvectors}), but without the factor of $1/2$.
After doing this, we obtain
\begin{equation}
\left( \begin{array}c 
  \dot a_x \\ \dot a_y \\ \dot a_\theta \\ \dot a_0
\end{array} \right) 
= k\mathbf{aSc}^T \mathbf{v}.
\label{findA}
\end{equation}
Eq.~(\ref{findA}) represents four equations, one corresponding to each
of $a_x$, $a_y$, $a_\theta$, and $a_0$.  The first three coefficients
are known from Eq.~(\ref{Fsolution}), thus the first three equations above
can be used to find the three components of $\mathbf{v}$.
Then the last equation can be used to determine $\dot a_0$.  
The displacement of the disk from its equilibrium position is
obtained by integrating $\mathbf{v}$.

\subsection{Application}
\label{MD}
\subsubsection{Shallow angles: $\tan\phi<\mu$}

When $\tan\phi<\mu$, the contacts remain non-sliding until the disk
begins to rotate.  Therefore, we can consider the entire experiment
to take place during one time interval.

When the contacts are non-sliding,
\begin{equation}
k\mathbf{aSc}^T = 2k \left( \begin{array}{ccc}
1 & 0 & r\cos\phi \\
0 & 1 & 0 \\
0 & 0 & r\sin\phi \\
0 & 0 & 0
\end{array} \right).
\label{nonsliding}
\end{equation}
Using this result in Eq.~(\ref{findA}), we have
\begin{eqnarray}
\dot a_x &=& 2kv_x + 2kr\omega\cos\phi,\cr
\dot a_y &=& 2kv_y,\cr
\dot a_\theta &=& 2kr\omega\sin\phi,\cr
\dot a_0 &=& 0.
\label{ans}
\end{eqnarray}
From the last line, we can see that $a_0$ remains constant as long
as the contacts are non-sliding.  This explains why $a_0=0$ for all
the MD simulations in Fig.~\ref{smallscan}: $a_0=0$ at the beginning
of the simulation, and thus remains so until the disk rolls.  This
also explains why $a_0$ is constant when $g$ is decreases in
Figs.~\ref{intermediate} and \ref{steep}.

Next let us calculate the disk's motion.
Eqs.~(\ref{ans}) can be solved for the velocities and integrated. 
In this way, the position $\mathbf{r}$ of the disk can be shown to be
\begin{equation}
\mathbf{r} = \frac{1}{2k}\left( \begin{array}c
0 \\ -mg \\ \tau/(r^2 \sin^2\phi)
\end{array} \right).
\end{equation}
Because no contact changes its status, the motion is perfectly
reversible.  This is another way to see that the packing has
no memory when $\tan\phi<\mu$.

\subsubsection{Intermediate slopes: $\mu < \tan\phi < 1/\mu$}

When $\tan\phi>\mu$ and $t<t_A$, the contacts are sliding with
$T_\alpha=-\mu R_\alpha$ and $T_\beta=\mu R_\beta$:
\begin{equation}
k\mathbf{aSc}^T = 2k\cos^2\phi \left( \begin{array}{ccc}
\tan^2\phi-\mu\tan\phi & 0 & 0 \\
0 & 1+\mu\tan\phi & 0 \\
-\tan\phi-\mu\tan^2\phi & 0 & 0 \\
0 & \tan\phi-\mu & 0
\end{array} \right).
\end{equation}

Changes in contact status will occur, so it is necessary to subdivide the
experiment into time intervals.  The first change of status occurs at $t=t_A$,
when the derivative of the gravity changes sign, so we define our
first interval to end at $t=t_A$.
We can find the particle displacements using the same method as 
before, and find,
\begin{equation}
r_y(t_A) = -\frac{mg_\mathrm{max}}{2k}\frac{\sec^2\phi}{1+\mu\tan\phi}.
\label{rdropslide}
\end{equation}
As before, $r_x(t_A)=r_\theta(t_A)=0$.
We also have $a_0(t_A) = a_\mathrm{mem}$, with
\begin{equation}
a_\mathrm{mem} = mg_\mathrm{max} \frac{\tan\phi-\mu}{1+\mu\tan\phi}.
\label{adropslide}
\end{equation}

When the gravitational force begins to decrease, all contacts are non-sliding.
The displacements must be calculated using the matrix given in
Eq.~(\ref{nonsliding}).  As before, the particle rises by a distance
$m(g_\mathrm{max}-g_*)/(2k)$, while $a_0$ remains unchanged.  The
particle's position is now
\begin{equation}
r_y = -\frac{mg_* + a_\mathrm{mem}}{2k}.
\label{heightdropslide}
\end{equation}
It is important to realize that this is \textsl{not} the same as would
have been obtained if the gravity was simply increased directly to
$g_*$.  This is another expression of the packing's memory, but it is
much more convenient to think of the memory being located in $a_0$.
Eq.~(\ref{heightdropslide}) mingles information about the past with
information about the present, whereas Eq.~(\ref{adropslide})
contains information
only about the past.  This is a consequence of using the basis
in Eq.~(\ref{CDvectors}), where what is determined directly by the imposed
forces is separated from what is not.

When the torque is applied, the contacts are still non-sliding, and
they remain so until one of the contacts becomes sliding.  We can
determine which contact slides first by checking if $a_\mathrm{mem}$ is greater
than or less than $a_2$.  If $a_\mathrm{mem}<a_2$, then the system meets
the lower side of the funnel first, meaning contact $\alpha$ slides.
Otherwise, it meets the upper edge first, and contact $\beta$ slides.
Using Eqs.~(\ref{aslip2}) and (\ref{adropslide}), we obtain that
$a_\mathrm{mem}>a_2$ is equivalent to
\begin{equation}
g_\mathrm{max} > g_*\left( 1 +
   \frac{\mu}{\tan\phi}\frac{1+\tan^2\phi}{1+\mu^2}\right).
\label{betawillslip}
\end{equation}
If this condition holds, then as $\tau$ increases,
$a_0$ decreases to maintain equality in Eq.~(\ref{beta+}).  Eventually,
equality is obtained in Eq.~(\ref{alpha+}), i.e., contact $\alpha$
begins to slip as well, and the disk rotates.  This occurs when
$a_0=a_2$, with $a_2$ given in Eq.~(\ref{aslip2}).
On the other hand, if Eq.~(\ref{betawillslip}) is not obeyed, than
contact $\alpha$ slips first.  Then $a_0$ increases until $a_0=a_2$, and
the disk begins to rotate.  Evaluating Eq.~(\ref{betawillslip}) for
$\mu=0.5$, $\phi=50^\circ$, one obtains $g_\mathrm{max} \approx 1.8 g_*$,
consistent with Fig.~\ref{intermediate}, where simulations with
$g_\mathrm{max}/g=2,4,6,8$ all meet the upper edge of the funnel.

\subsubsection{Steep slopes: $\tan\phi>1/\mu$}
\label{steepMDtheory}

For $\tan\phi>1/\mu$, all the calculation in the previous section
applies.  The only difference is that the slope of the line given
by equality in Eq.~(\ref{beta+}) changes sign.  This line is shown
in Fig.~\ref{steep}, and if the system were
to behave as before, it would follow this line upward indefinitely,
leading to an infinite torque.
However, the simulations show otherwise.
When Eq.~(\ref{beta+}) is attained, the system suddenly jumps to
the rotating state, in a behavior qualitatively different from the
other cases.  This jump occurs when $T_\beta = - \mu R_\beta$,
so this suggests that we ought to calculate the displacements
of the disk in this situation.
We calculate the matrix $k\mathbf{aSc}^T$ as before, and
assume that the torque is increasing.
Calculating the velocity of the disk $\mathbf{v}$ in the usual way, one obtains
\begin{equation}
\mathbf{v} = \frac{\dot\tau}{2k} \frac{\csc\phi}{(1-\mu\tan\phi)}
\left( \begin{array}c
\mu - \tan\phi + \sec\phi\csc\phi\\
\tan\phi(\mu+\tan\phi) \\
\frac{1}{r} \sec^2\phi\csc\phi
\end{array} \right).
\label{vweird}
\end{equation}
Note the factor $1-\mu\tan\phi$ in the denominator, which is positive
when $\tan\phi<1/\mu$ and negative when $\tan\phi>1/\mu$.  
Let us pause for a moment to consider what this change of sign means
for $\omega$.  Applying a positive torque to the disk means trying to
rotate it in the positive direction, i.e., trying to impose $\omega > 0$.
When $\tan\phi<1/\mu$, Eq.~(\ref{vweird}) shows that rotating the particle
in this direction generates contact forces that balance an increasing torque.
Therefore, the configuration is stable: when $\tau$ increases, the particle
rotates in the positive direction, and the contact forces balance the increased
torque.  On the other hand, when $\tan\phi>1/\mu$, the particle must be
rotated in the \textsl{negative} direction to balance an increasing torque.
But such a torque will always rotate the disk in the positive direction.
Thus, the configuration is unstable.  The small increase in torque causes
the particle to rotate in the positive direction as before.  But this rotation
causes the forces opposing the torque to decrease, which in turn causes the
particle to rotate farther in the positive direction.  This leads to a rapid
change in the contact forces as the particle begins to rotate.
Another way to see this is to calculate $\mathbf{v}^T \mathbf{Qv}$ using
the value of $\mathbf{v}$ given above. One obtains
\begin{equation}
\mathbf{v}^T \mathbf{Qv} = 
\frac{\dot\tau^2}{2kr} \frac{\csc^2\phi \sec^2 \phi}{1-\mu\tan\phi},
\end{equation}
which becomes negative for $\tan\phi>1/\mu$.  Thus the particle interactions
no longer oppose the applied force, but amplify it.  This leads
to an exponential growth of the motion, and the rapid change
in the forces observed in the simulations.

We can now predict the yielding torque in the MD simulations.
If Eq.~(\ref{betawillslip}) is not satisfied, the system meets the
lower edge of the funnel that functions normally, and the disk rotates
when $\tau=\tau_2$.  An example of this is the MD simulation at 
$g_\mathrm{max}/g=2$ in Fig.~\ref{steep}.  On the other hand, if
Eq.~(\ref{betawillslip}) is satisfied, the system meets the other
branch of the funnel, and immediately starts to rotate.  We can calculate
the torque where this happens because the value of $a_0$ is known.
Thus inserting the value of $a_0$ given in Eq.~(\ref{adropslide}) 
into Eq.~(\ref{beta+}) and solving for the torque gives
\begin{equation}
\tau = mr \sin\phi
\left[g_* + g_\mathrm{max}\left(\frac{\tan\phi-\mu}{\tan\phi+\mu}
   \frac{\mu\tan\phi-1}{\mu\tan\phi+1} \right) \right].
\end{equation}
This result accurately predicts the yielding torque, as shown in
Fig.~\ref{scan}.

\section{Conclusions}

We have carried out an intensive investigation of a very simple
system.  This has enabled us to precisely understand certain 
principles of the quasi-static motion of granular packings.  First of
all, there is the relation between memory and indeterminacy.  When particles
are assumed to be rigid, the forces cannot be uniquely determined.
A certain combination of forces, whose amplitude is given by the
coefficient $a_0$ in Eq.~(\ref{Fexpand}), causes no acceleration, 
and thus cannot be deduced by examining the equations of static equilibrium.  
However, its value
can be constrained by enforcing Coulomb friction and the absence of cohesion
at all the contacts.  On the other hand, when the particles
are assumed to be deformable, $a_0$ contains
information about the history of the packing.
This finding suggests that methods that average over different possible
force networks \cite{FNensemble} are really averaging over past
histories of the packing.

Secondly, we were able to precisely understand the connection between
indeterminacy and motion.  Depending on the angle $\phi$, the onset
of motion of the disk can take two different forms.  When $\tan\phi<1/\mu$,
the range of allowed values of $a_0$ decreases continuously,
and vanishes precisely at the onset of motion, and
the torque needed to turn the disk is independent of method and history
of the packing.  Furthermore, the forces change slowly and continuously.
A second mode of failure is more dramatic.  When $\tan\phi>1/\mu$,
the indeterminacy is infinite, i.e., there is no upper bound on $a_0$.
At the onset of motion, the forces change discontinuously (on the time
scale of changes in the imposed torque), and the torque needed
to turn the disk depends on the history and method.  This mode of
failure is related to a curious situation in which the contact forces
no longer oppose the applied forces, but enhance it.

This work shows that giving the CD iterative solver the
the solution of the last time step as the initial guess is not just
a numerical trick to accelerate convergence.  It is an essential part
of the method that represents a real physical process at work in
granular material, namely the memory of the packing encoded in
the combinations of force that cause no acceleration.  In certain situations,
this memory could have an important effect on the behavior of
the packing.

It is hoped that the findings presented in this paper will deepen
and clarify our understanding of the quasi-static deformation of
much larger packings.  When the number of particles is larger,
there is not just one undetermined combination of forces,
but many.  However, one can expect that the effects studied here
will be present in these more complicated situations, perhaps
combined with each other in new and unexpected ways.

\begin{acknowledgements}
The authors want to acknowledge the EU project \textit{Degradation and Instabilities
in Geomaterials with Application to Hazard Mitigation} (DIGA) in the
framework of the Human Potential Program, Research Training Networks
(HPRN-CT-2002-00220).  The authors wish to thank F. Alonso-Marroqu\'\i n
and J.J. Moreau for encouragement and interesting discussions.
\end{acknowledgements}

\bibliographystyle{prsty}

\end{document}